\documentclass[twocolumn]{aastex62}
\usepackage{float,graphicx,amsmath,multirow}
\usepackage{color}
\usepackage[version=3]{mhchem}
\usepackage{booktabs}
\usepackage{dcolumn}
\newcommand{\nth}{$^{\mbox{th}}$}

\begin{document}
\title{Interstellar Detection of 2-Cyanocyclopentadiene, \ce{C5H5CN}, a Second Five-Membered Ring Toward TMC-1}
\author{Kin Long Kelvin Lee}
\affiliation{Department of Chemistry, Massachusetts Institute of Technology, Cambridge, MA 02139, USA}
\author{P. Bryan Changala}
\affiliation{Center for Astrophysics $\vert$ Harvard \& Smithsonian, Cambridge, MA 02138, USA}
\author{Ryan A. Loomis}
\affiliation{National Radio Astronomy Observatory, Charlottesville, VA 22903, USA}
\author{Andrew M. Burkhardt}
\affiliation{Center for Astrophysics $\vert$ Harvard \& Smithsonian, Cambridge, MA 02138, USA}
\author{Ci Xue}
\affiliation{Department of Chemistry, University of Virginia, Charlottesville, VA 22904, USA}
\author{Martin A. Cordiner}
\affiliation{Astrochemistry Laboratory and the Goddard Center for Astrobiology, NASA Goddard Space Flight Center, Greenbelt, MD 20771, USA}
\affiliation{Institute for Astrophysics and Computational Sciences, The Catholic University of America, Washington, DC 20064, USA}
\author{Steven B. Charnley}
\affiliation{Astrochemistry Laboratory and the Goddard Center for Astrobiology, NASA Goddard Space Flight Center, Greenbelt, MD 20771, USA}
\author{Michael C. McCarthy}
\affiliation{Center for Astrophysics $\vert$ Harvard \& Smithsonian, Cambridge, MA 02138, USA}
\author{Brett A. McGuire}
\affiliation{Department of Chemistry, Massachusetts Institute of Technology, Cambridge, MA 02139, USA}
\affiliation{National Radio Astronomy Observatory, Charlottesville, VA 22903, USA}
\affiliation{Center for Astrophysics $\vert$ Harvard \& Smithsonian, Cambridge, MA 02138, USA}

\correspondingauthor{Kin Long Kelvin Lee, Brett A. McGuire}
\email{kelvlee@mit.edu, brettmc@mit.edu}

\begin{abstract}

Using radio observations with the Green Bank Telescope, evidence has now been found for a second five-membered ring in the dense cloud Taurus Molecular Cloud-1 (TMC-1).  Based on additional observations of an ongoing, large-scale, high-sensitivity spectral line survey (GOTHAM) at centimeter wavelengths toward this source, we have used a combination of spectral stacking, Markov chain Monte Carlo (MCMC), and matched filtering techniques to detect 2-cyanocyclopentadiene, a low-lying isomer of 1-cyanocyclopentadiene, which was recently discovered there by the same methods.  The new observational data also yields a considerably improved detection significance for the more stable isomer and evidence for several individual transitions between 23--32\,GHz. Through our MCMC analysis, we derive cospatial, total column densities of $8.3\times10^{11}$ and $1.9\times10^{11}$\,cm$^{-2}$ for 1- and 2-cyanocyclopentadiene respectively, corresponding to a ratio of ${\sim}$4.4 favoring the former. The derived abundance ratios point towards a common formation pathway---most likely being cyanation of cyclopentadiene by analogy to benzonitrile.

\end{abstract}


\section{Introduction}

\begin{figure}
    \centering
    \includegraphics[width=\columnwidth]{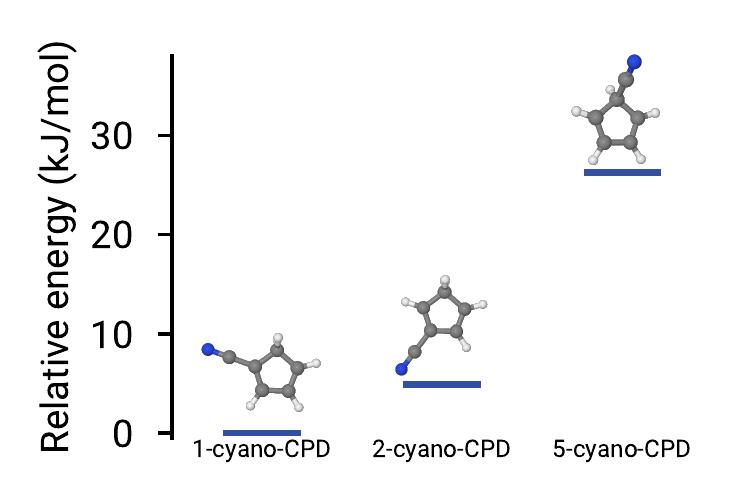}
    \caption{Geometric structures of the three low-lying cyanocyclopentadiene isomers, 1-, 2-, and 5-cyanocyclopentadiene, $c$-\ce{C5H5CN} and G3//B3LYP \citep{baboul_gaussian-3_1999} energetics at 0\,K.  The dipole moment projections are $\mu_a$ = 4.15\,D and $\mu_b$ = 0.27\,D for 1-cyano-CPD, and $\mu_a$ = 4.36\,D and $\mu_b$ = 0.77\,D for 2-cyano-CPD~\citep{sakaizumi:3903}. 5-cyano-CPD has not yet been observed, but possesses similarly favorable dipole moments ($\mu_a$ = 3.91\,D, $\mu_b$ = 1.47\,D calculated at the $\omega$B97X-D/cc-pVQZ level of theory).
    }
    \label{molecules}
\end{figure}

The recent astronomical detection of 1-cyano-1,3-cyclopentadiene (Fig.~\ref{molecules}; hereafter 1-cyano-CPD) and other CN-functionalized ring molecules in the starless cloud core, Taurus Molecular Cloud-1 (TMC-1), has opened up an entirely new and unexplored area of aromatic organic chemistry in space.  In rapid succession, evidence has been found for both five-membered~\citep{mccarthy_interstellar_2020} and six-membered rings~\citep{mcguire:202}, and even bicyclic ones~\citep{mcguire:submitted} in a primordial gas cloud which has long been known to be exceedingly rich in highly unsaturated carbon chains, most notably cyanopolyynes and acetylenic free radicals, among others.  Intriguingly, the derived abundances of these rings exceed---in some cases by many orders of magnitude---those predicted from chemical models which well reproduce the abundance of a wide assortment of chains regardless of length.  For this reason, questions as to the relative importance of bottom-up formation pathways versus inheritance from previous top-down routes that might survive to the dense cloud phase have been raised, but remain poorly constrained at present.

Substitution of a H atom of cyclopentadiene, $c$-\ce{C5H6} with a nitrile group yields three possible cyanocyclopentadienes.  Quantum chemical calculations by \citet{mccarthy_interstellar_2020}, shown in Figure \ref{molecules}, predict 1-cyano-CPD is the most stable isomer, followed closely by 2-cyano-1,3-cyclopentadiene (2-cyano-CPD; 5\,kJ/mol or 600\,K) and then 5-cyano-1,3-cyclopentadiene, which lies far higher in energy (by 26\,kJ/mol or ${\sim}$3130\,K).  These findings are in agreement with those from an earlier study \citep{wentrup:4375} and predicted from conjugation arguments.

The rotational spectra of both 1- and 2-cyano-CPD have been reported by several groups \citep{ford:326,sakaizumi:3903}, with the most recent set of measurements \citep{mccarthy:5170,mccarthy_interstellar_2020} yielding very sharp lines ($\sim$5\,kHz FWHM linewidths) and rest frequencies accurate to 2\,kHz or better between 7 and 30\,GHz. In terms of equivalent radial velocity, this accuracy corresponds to  $<0.1$\,km s$^{-1}$, which is more than sufficient to conduct a rigorous search in the coldest, most quiescent molecular clouds.  Both isomers are highly polar, with measured dipole  moments  along  their $a$-inertial  axes in excess of 4\,D [4.15(15)\,D~and 4.36(25)\,D, respectively; \citep{sakaizumi:3903}], and comparable to that of benzonitrile ($\mu_a=4.5152$\,D; \citep{wolhfart:119}).

On the basis of highly accurate laboratory rest frequencies, and the first data release (DR1) from a large-scale, high sensitivity spectral line survey, GBT Observations of TMC-1: Hunting for Aromatic Molecules (GOTHAM), predominately in the K (18--27\,GHz) and K$_a$ bands (26--40\,GHz), a number of the authors here recently reported the astronomical detection of 1-cyano-CPD using spectral stacking and matched filtering techniques \citep{mccarthy_interstellar_2020}.  From these observations, which represent ${\sim}$30\% of the project goal, an upper limit for 2-cyano-CPD relative to 1-cyano-CPD was estimated to be roughly 1/3. Because the abundance ratio in the laboratory ranges from 1/2 to 1/4 \citep{sakaizumi:3903,mccarthy:5170}, it was unclear if 1-cyano-CPD is formed selectively in TMC-1 or if the apparent absence of  2-cyano-CPD is simply a question of sensitivity, given its somewhat lower stability and consequently lower abundance.  With the second data release (DR2) and additional laboratory measurements, this ambiguity has now been resolved, and in doing so a common formation for this isomeric pair is implicated.


\section{Observations and Data Analysis}

The observations in DR2 fold in new observations made between February 2018 and June 2020 on the Robert C. Byrd 100-m Green Bank Telescope in Green Bank, West Virginia under project codes GBT18A-333, GBT18B-007, and GBT19A-047.  Although the frequency range of the present dataset is only slightly wider than that covered in DR1 \citep{mcguire:l10}, it is considerably more sensitive at some frequencies \citep{mcguire:submitted}. The spectral coverage of DR2 extends from 7.906 to 33.527 GHz (25.6 GHz bandwidth) with continuous coverage between 22--33.5 GHz, at a uniform frequency resolution of 1.4\,kHz (0.05--0.01\,km/s in velocity) and an RMS noise of ${\sim}2$--20\,mK ($T_A*$) across the spectrum.

As before, the target was the cyanopolyyne peak of TMC-1 at (J2000) $\alpha$~=~04$^h$41$^m$42.50$^s$ $\delta$~=~+25$^{\circ}$41$^{\prime}$26.8$^{\prime\prime}$.  The calibrator source for pointing and focus observations was J0530+1331; focus and pointing offsets were performed at the beginning of each observing session, and subsequently every 1 to 2\,h, depending on the weather; typical pointing convergence was $\lesssim$5$^{\prime\prime}$.  Observations were performed using position switching (ON-OFF), in which the target and the off position were observed in a sequential manner, each for 2\,min.  The off position was chosen to be 1$^{\circ}$ off target and was confirmed to be clear of emission.  Additionally data from project GBT17A-164 and GBT17A-434  have also been folded in the DR2.  The observing strategy for this archival data is outlined in \cite{mcguire:202}, but it is very similar to that used here.  To ensure uniformity and consistency with the present data set, the archival data were re-calibrated and re-reduced.   Uncertainty due to flux calibration is expected to be $\sim$20\%, based on complementary VLA observations of the flux-calibrator source J0530+1331 \citep{mcguire_early_2020}.

A detailed description of the data analysis and statistical methods, including the procedure for spectral stacking and matched filtering, are presented elsewhere \citep{Loomis:2020aa}, so only a brief summary is provided here. Briefly, many small frequency regions, each centered around a predicted transition frequency of a target species, are extracted from the full dataset of observations. Any window containing an obvious spectral feature (e.g., $>$5$\sigma$) is omitted so as to avoid any interlopers: this corresponds to 275 transitions for 1-cyano-CPD and 326 for 2-cyano-CPD, without interlopers detected. A signal-to-noise weighted average of the spectra was then performed based on the expected intensity of the line  and the RMS noise of the observations.  Only transitions of the target species that have a predicted flux $\geq 5$\% of the strongest line are considered in our analysis. This procedure is built into the \textsc{molsim} package \citep{lee_molsim_2020}, which performs the spectral simulation and wraps the affine-invariant MCMC implementation of \textsc{emcee} \citep{foreman-mackey_emcee_2013} and posterior analysis routines from \textsc{ArviZ} \citep{kumar_arviz_2019}.

\begin{table*}[tbh!]
\centering
\caption{Spectroscopic constants of 1-cyano-CPD and 2-cyano-CPD. The fits were performed with the Watson A-reduced Hamiltonian including quartic centrifugal distortion. All values are given in MHz with 1$\sigma$ uncertainties in parentheses. Values bracketed with [~] were held fixed.}
\label{table:labdata}

3
\newcolumntype{.}{D{.}{.}{-6}}
\begin{tabular}{c|.....}
\toprule
\multirow{2}{*}{Parameter} & \multicolumn{2}{c}{1-cyano-CPD} && \multicolumn{2}{c}{2-cyano-CPD} \\ 
\cline{2-3} 
\cline{5-6}
\multicolumn{1}{c}{} & \multicolumn{1}{c}{This work} & \multicolumn{1}{c}{\cite{mccarthy_interstellar_2020}} && \multicolumn{1}{c}{This work} & \multicolumn{1}{c}{\cite{mccarthy_interstellar_2020}} \\
\midrule
$A$ &8352.981(10)&8352.98(2)&&8235.592(14)&8235.66(4)\\
$B$ &1904.2522(2)&1904.2514(3)&&1902.0748(3)&1902.0718(2)\\
$C$ &1565.3652(2)&1565.3659(3)&&1559.6472(2)&1559.6502(2)\\ 
$\Delta_J \times 10^3$ &0.0743(11)&0.0701(15)&&0.0686(11)&0.0561(37)\\
$\Delta_{JK} \times 10^3$ &2.354(8)&2.361(15)&&2.287(21)&2.286(46)\\
$\Delta_K \times 10^3$ &[0.17561]^a&[0.]&&[0.32391]^a&[0.]\\
$\delta_J \times 10^3$ &0.0133(5)&0.0120(11)&&0.0134(6)&[0.]\\
$\delta_K \times 10^3$ &1.48(9)&1.21(13)&&1.10(9)&[0.]\\
$\chi_{aa}(\mathrm{N})$ &-4.1810(11)&-4.1796(21)&&-4.2429(13)&-4.234(6)\\
$\chi_{bb}(\mathrm{N})$ &2.3016(14)&2.3052(26)&&2.2475(16)&2.236(7)\\
\midrule 
$N_\text{lines}$~$^b$ &154&68&&110&38\\
$(J, K_{a})_\text{max}$ &(11,3)&(9,3)&&(10,2)&(5,2)\\
\bottomrule
\multicolumn{6}{l}{$^a$ Calculated at the $\omega$B97X-D/cc-pVQZ level of theory.}\\
\multicolumn{6}{l}{$^b$ The number of hyperfine-resolved transitions included in the fit.}
\end{tabular}
\end{table*}
 
Given that the 1-cyano-CPD isomer was characterized in our earlier work, we used the posterior distributions obtained in \citet{mccarthy_interstellar_2020} as priors for both 1-cyano-CPD and 2-cyano-CPD MCMC simulations here---effectively refining the previous model with the new experimental and observational data. The model space comprises 14 parameters corresponding to four known velocity components within TMC-1 \citep{Dobashi:2018kd,Dobashi:2019ev}, each having an independent source size [SS], column density ([$N_T$], and radial velocity [$v_{lsr}$], while a common excitation temperature [$T_{ex}$] and linewidth [$dv$] are assumed. The parameters are used to simulate the expected flux in a forward model, taking into account beam dilution and optical depth effects, with the MCMC sampling guided by the observed spectra. Finally, to determine an overall significance of a detection, the model spectra are stacked using identical weights, and that stacked model is then used as a matched filter that is cross-correlated with the stacked observations.  The resulting response spectrum provides a lower limit on the statistical significance to the detections.
 

\section{Laboratory Measurements}

The astronomical detection of 1-cyano-CPD was based on laboratory measurements made using a cavity-enhanced Fourier transform microwave spectrometer in which this molecule and many others \citep{mccarthy:5170,mccarthy_interstellar_2020} were produced in discharge of benzene and molecular nitrogen. To improve the accuracy of the spectroscopic constants of 1-cyano-CPD and measure a comparable number of transitions for 2-cyano-CPD, a discharge of dicyclopentadiene (the Diels-Alder dimer of CPD) and acetonitrile was used here.  This precursor combination yielded nearly a ten-fold increase in line intensity for both isomers relative to our earlier work. For 1-cyano-CPD, the number of hyperfine components in the laboratory data set approximately doubled, while a three-fold increase was achieved for 2-cyano-CPD.  In doing so, the frequency range of the measurements increased commensurately: from 30 to 36\,GHz for 1-cyano-CPD, and from 19\,GHz to 33\,GHz for  2-cyano-CPD. The best-fit spectroscopic constants for both species---the most complete summary of the rotational data yet---are summarized in Table~\ref{table:labdata}.

As part of these measurements, improved SNR for both isomers enabled transitions between higher $J$ and $K_a$ levels to be observed. Inclusion of these weaker transitions proportionally increases the rotational partition functions, and alters somewhat the column densities derived for both species. At 300\,K, the new values for $Q_\mathrm{rot}$ are approximately 1.4 times larger than in \cite{mccarthy_interstellar_2020}, resulting in a substantially lower column density, and one outside the 2$\sigma$ uncertainties reported in our previous analysis.


\begin{figure*}[tbh!t!]
    \centering
    \includegraphics{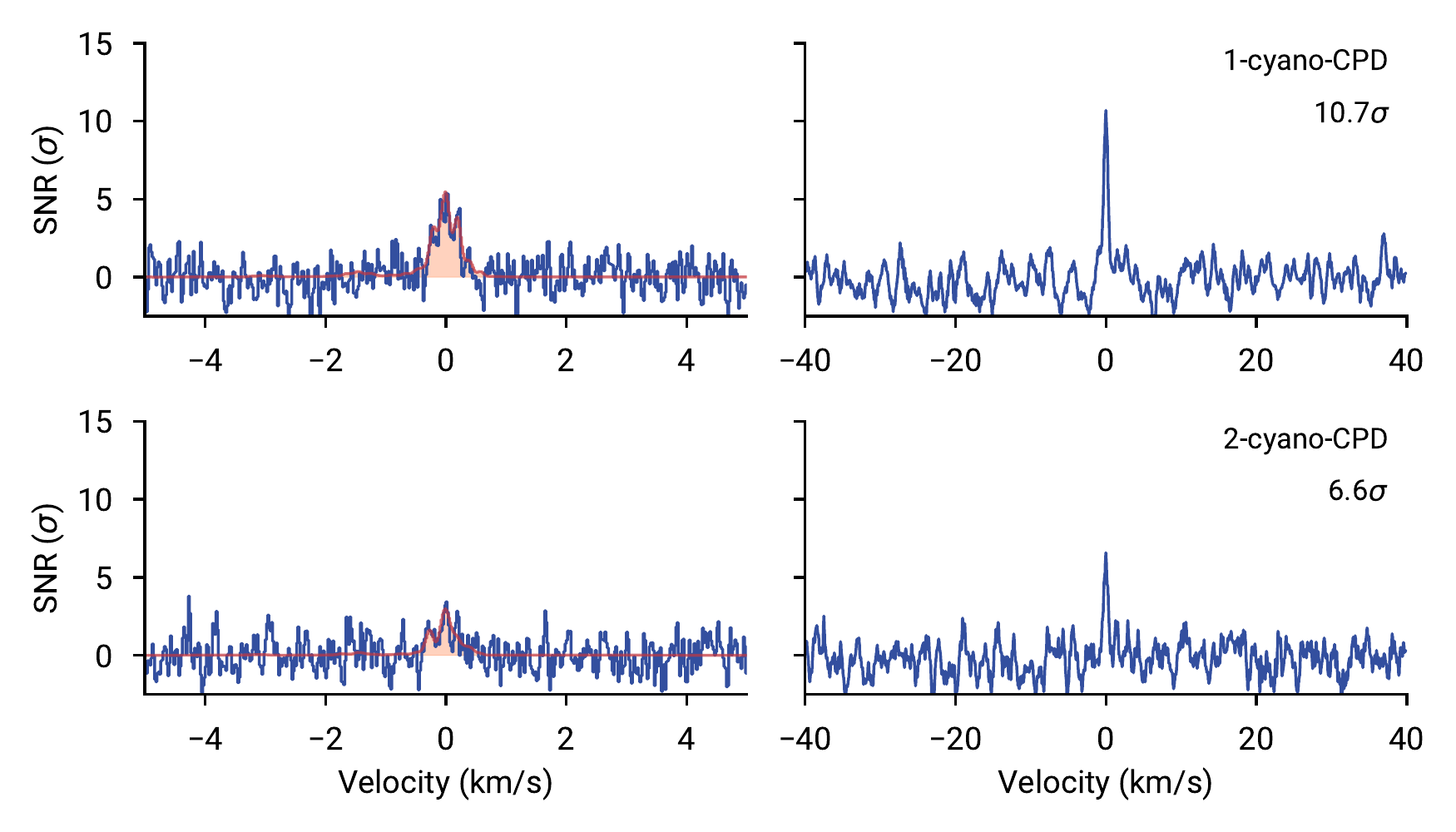}
    \caption{Velocity stacked and matched filter spectra of 1-cyano-CPD and 2-cyano-CPD. (\emph{Left}) The stacked spectra from the GOTHAM DR2 data are display in blue, overlaid with the expected line profile in red from our MCMC fit to the data.  The signal-to-noise ratio is on a per-channel basis. (\emph{Right}) Impulse response functions of the stacked spectra of same molecules using the simulated line profiles as matched filters. The intensity scales are the signal-to-noise ratios of the response functions when centered at a given velocity. The values denoted in each figure indicate the peak of the impulse response functions which provide a minimum significance for the detection.}
    \label{fig:stacks}
\end{figure*}

\section{Results and Discussion}

\begin{figure*}[hbt!]
    \centering
    \includegraphics{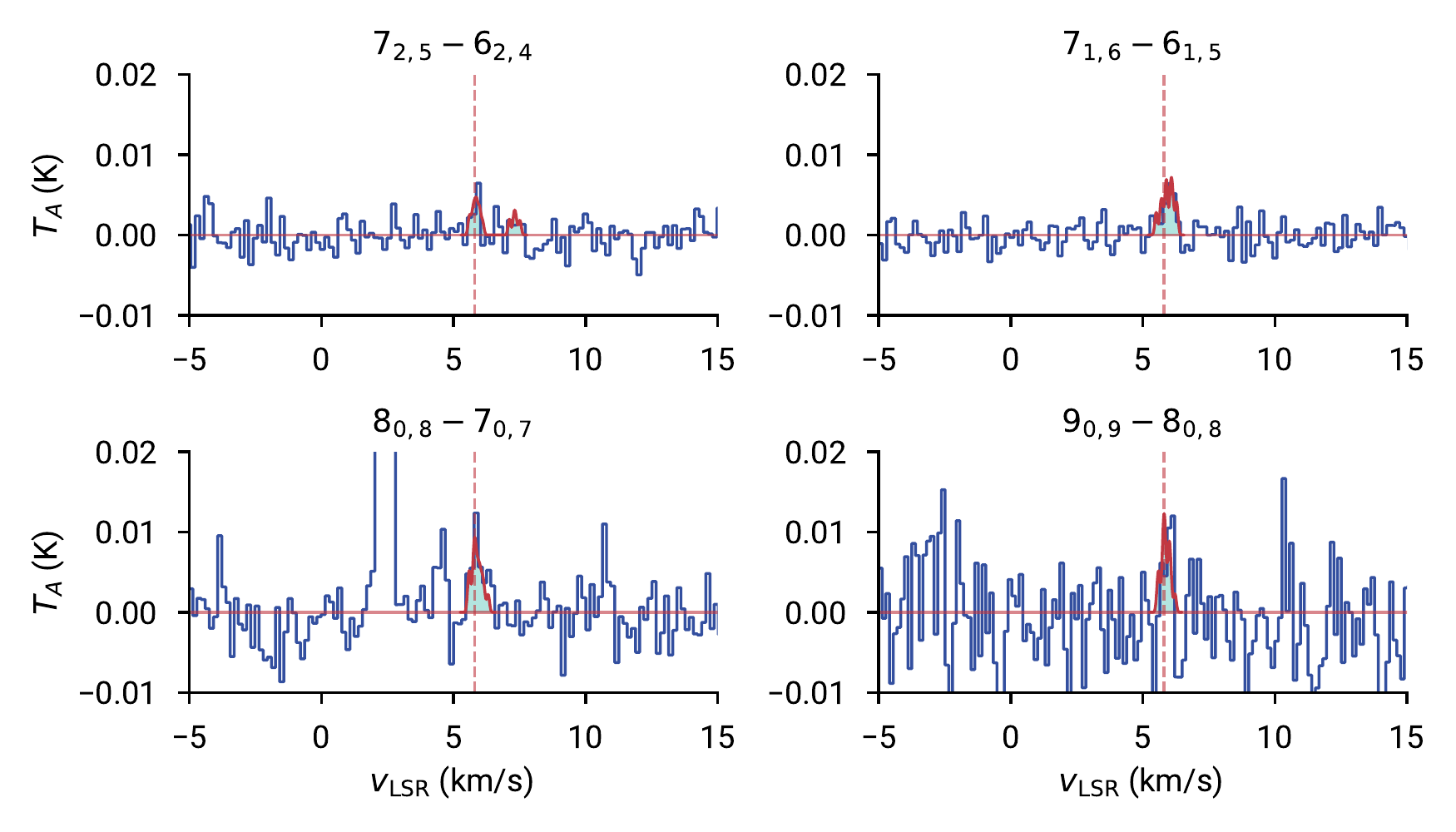}
    \caption{Four rotational transitions of 1-cyano-CPD observed towards TMC-1 using the 100-m GBT telescope.  The overlapping red trace and shaded regions represent the simulated flux based on mean parameters derived from the MCMC posterior. The corresponding asymmetric top ($J_{K_a,K_c}$) quantum numbers for each transition are shown above each feature. The dashed line indicates the nominal 5.8 km/s source velocity.}
    \label{fig:lines}
\end{figure*}

Shown in Fig.~\ref{fig:stacks} are stacked spectra and the impulse responses for the two  cyanocyclopentadiene isomers that have been derived from  the DR2 data. Additional results of the MCMC results can be found in the Appendix. Relative to the DR1 results, these additional observations improve the detection significance of 1-cyano-CPD by a factor of 1.8 (from 5.8$\sigma$ to 10.7$\sigma$), and more importantly, provide compelling evidence for 2-cyano-CPD.  As the visual representations of the statistical analysis illustrate, we can now conclude with good confidence that both isomers are present in TMC-1. Furthermore, albeit at low signal-to-noise ratio (SNR), evidence for several individual rotational lines of 1-cyano-CPD as shown in Fig \ref{fig:lines}; these correspond to the strongest transitions at ${\sim}8$\,K in regions of our spectra with varying SNR.

Tables~\ref{table:1cyano} and \ref{table:2cyano} summarize the best-fit values from our MCMC analysis.  In agreement with prior work \citep{Dobashi:2018kd,Dobashi:2019ev}, four distinct velocity components in the source with nominally independent four column densities and four source sizes were included in our model. In comparison with our DR1 analysis, the higher frequency coverage in DR2 provides better constraints on the source sizes and in turn the derived column densities. Most of the derived quantities are similar in magnitude to those reported previously, and, as before, the source sizes remain poorly constrained.  Most importantly, an abundance ratio of $4.4(6)$ (1$\sigma$ uncertainty) has been determined for the two isomers; we note, however that this value treats the two models as statistically independent, and a proper estimate would require explicit treatment of the column density covariances between the two molecules.

The astronomical detection of 2-cyano-CPD with a marginally similar abundance to 1-cyano-CPD strongly suggests both isomers are formed via a common pathway.  The simplest and most direct route involves the reaction of cyclopentadiene with CN radical.  Although apparently not  examined experimentally, theoretical calculations \citep{mccarthy_interstellar_2020} at the G3//B3LYP level of theory predict both 1-cyano-CPD and 2-cyano-CPD are formed exothermically and barrierlessly, irrespective of the small difference in their stabilities.  This reaction is analogous to benzonitrile formation from benzene and CN, which has been extensively studied by multiple experimental ~\citep{balucani:2000nw,trevitt_reactions_2010,lee_gas-phase_2019,Cooke:2020we} and theoretical techniques~\citep{Woon:2006ce,lee_gas-phase_2019}, and is known to occur rapidly at low temperature, implying the possibility of \textit{in situ} formation of cyanide derivatives of hydrocarbons under the cold, dark conditions of TMC-1. By analogy, the same logic implies the presence of cyclopentadiene in TMC-1, although the small permanent dipole moment [$\mu_b=0.420$\,D;~\citep{laurie:635,damiani:265}] of this hydrocarbon ring will make its direct detection very challenging in the radio band. It is also worth noting the abundance of the two cyanocyclopentadienes taken together close to that of benzonitrile [$1.60 \times 10^{12}$\,cm$^{-2}$; \cite{burkhardt:inpress}].  Since  both the five and six-membered rings likely form via reaction with a common precursor, the \ce{CN} radical, and the cyanation reactions are highly exothermic and barrierless, this ratio may reflect the nascent hydrocarbon abundances in TMC-1. Although there is no obvious \textit{a priori} expectation of the benzene/cyclopentadiene abundance ratio, the former is aromatic while the latter is not, and thus might be expected to more abundant based on thermodynamic stability.

In a more general sense, the present work further highlights that carbon chemistry of considerable richness and complexity lies just below the noise floor of previous spectral line surveys toward this well-known primordial molecular cloud.  Despite  discoveries of monocyclic and even bicyclic rings in rapid succession, this work has raised many more questions than it has helped answer.  Nevertheless, it serves to illustrate that there is potentially much more to be learned about a well-studied and until recently a seemingly well-understood source. In this context,  it would be disappointing if still other functionalized rings and potentially their precursors are not eventually found with sustained effort.

\section{Data access \& code}

Data used for the MCMC analysis can be found in the DataVerse entry \citep{DVN/K9HRCK_2020}. The code used to perform the analysis is part of the \textsc{molsim} open-source package; an archival version of the code can be accessed at \cite{lee_molsim_2020}.

\acknowledgments

M.C.M, K.L.K.L., and P.B.C. acknowledge financial support from NSF grants AST-1908576, AST-1615847, and NASA grant 80NSSC18K0396.  A.M.B. acknowledges support from the Smithsonian Institution as a Submillimeter Array (SMA) Fellow. S.B.C. and M.A.C. were supported by the NASA Astrobiology Institute through the Goddard Center for Astrobiology. C.X. is a Grote Reber Fellow, and support for this work was provided by the NSF through the Grote Reber Fellowship Program administered by Associated Universities, Inc./National Radio Astronomy Observatory and the Virginia Space Grant Consortium. The National Radio Astronomy Observatory is a facility of the National Science Foundation operated under cooperative agreement by Associated Universities, Inc.  The Green Bank Observatory is a facility of the National Science Foundation operated under cooperative agreement by Associated Universities, Inc.



\bibliography{mcm_0831_20,mccarthy_refs,new_bib,modeling,refs}
\bibliographystyle{aasjournal}

\appendix

\renewcommand{\thefigure}{A\arabic{figure}}
\renewcommand{\thetable}{A\arabic{table}}
\renewcommand{\theequation}{A\arabic{equation}}
\setcounter{figure}{0}
\setcounter{table}{0}
\setcounter{equation}{0}

\section{MCMC results}

The following summarizes the posteriors obtained for 1-cyano-CPD and 2-cyano-CPD from the MCMC modeling. The data comprises corner plots and statistics that visualizes and quantifies the sampling results. Corner plots are interpreted in two ways: the off-diagonal contours show the covariance between model parameters, while the diagonal traces are empirical cumulative distribution function (ECDF) plots. The latter indicates the integrated marginalized likelihood of a given parameter---in other words, the parameter space comprised by the posterior.

\subsection{1-cyano-CPD}

\begin{figure}[h!]
    \centering
    \includegraphics[width=\textwidth]{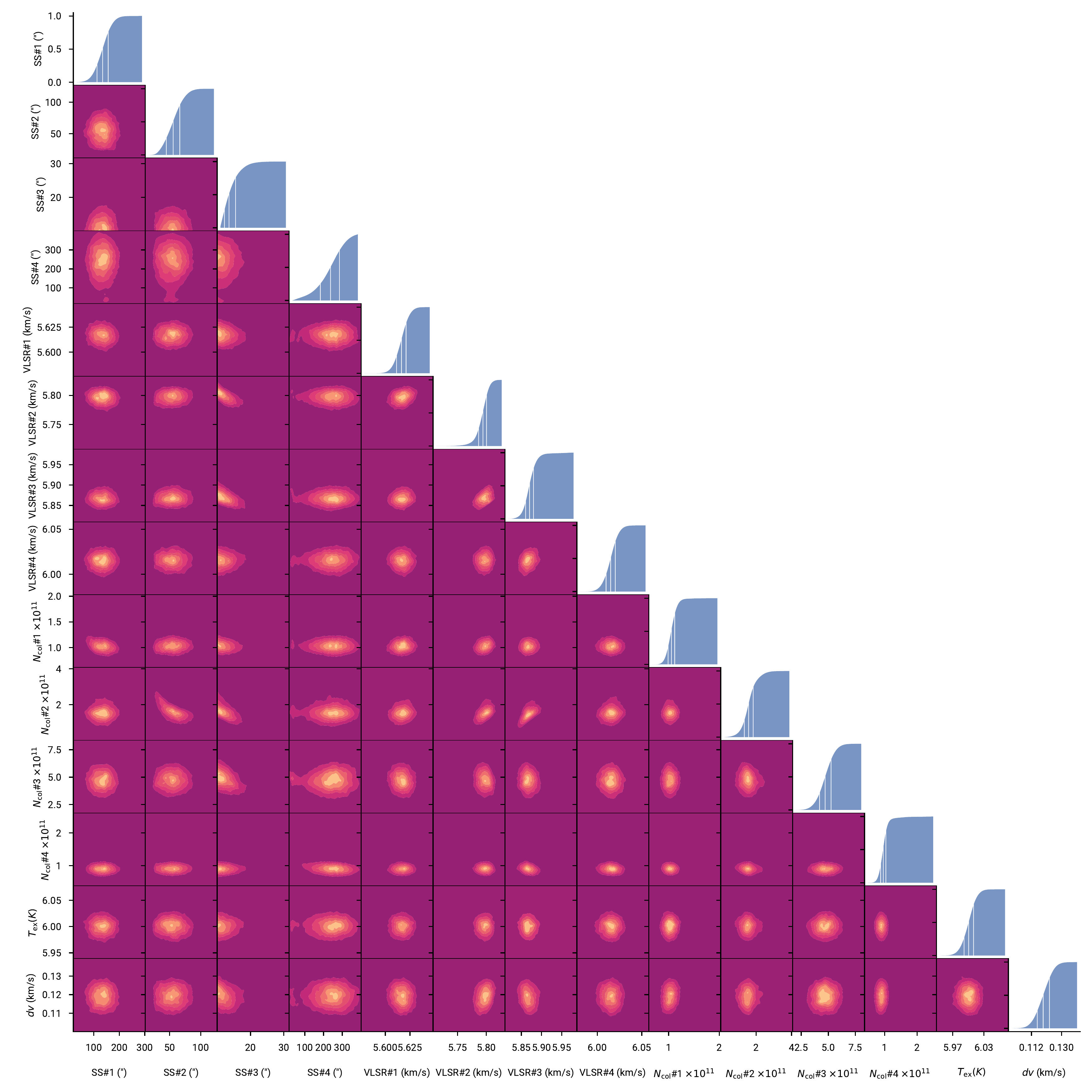}
    \caption{Corner plot for 1-cyano-CPD The diagonal traces correspond to ECDF plots, and off-diagonal plots show the kernel density covariance between model parameters. In the former, lines represent the 25{\nth}, 50{\nth}, and 75{\nth} percentiles respectively. The length scale for the kernel density plots is chosen with Scott's rule.}
    \label{fig:uno-corner}
\end{figure}

Figure \ref{fig:uno-corner} shows the posterior traces for the MCMC modeling analysis of 1-cyano-CPD. These simulations indicate detections in all four components, with the source sizes very poorly constrained and covariant with the column densities. Table \ref{table:1cyano} provides summary statistics of the marginalized posteriors.

\begin{table*}[hbt!]
\centering
\caption{1-cyano-CPD best-fit parameters from MCMC analysis.}
\label{table:1cyano}
\begin{tabular}{c c c c c c}
\toprule
\multirow{2}{*}{Component}&	$v_{lsr}$					&	Size					&	\multicolumn{1}{c}{$N_T^\dagger$}					&	$T_{ex}$							&	$\Delta V$		\\
			&	(km s$^{-1}$)				&	($^{\prime\prime}$)		&	\multicolumn{1}{c}{(10$^{11}$ cm$^{-2}$)}		&	(K)								&	(km s$^{-1}$)	\\
\midrule
\hspace{0.1em}\vspace{-0.5em}\\
C1	&	$5.617^{+0.015}_{-0.015}$	&	$132^{+67}_{-66}$	&	$1.03^{+0.20}_{-0.20}$	&	\multirow{6}{*}{$6.00^{+0.03}_{-0.03}$}	&	\multirow{6}{*}{$0.119^{+0.009}_{-0.010}$}\\
\hspace{0.1em}\vspace{-0.5em}\\
C2	&	$5.795^{+0.022}_{-0.023}$	&	$55^{+30}_{-30}$	&	$1.57^{+1.03}_{-0.86}$	&	&	\\
\hspace{0.1em}\vspace{-0.5em}\\
C3	&	$5.869^{+0.033}_{-0.033}$	&	$14^{+5}_{-4}$	&	$4.69^{+1.60}_{-1.62}$	&	&	\\
\hspace{0.1em}\vspace{-0.5em}\\
C4	&	$6.015^{+0.016}_{-0.017}$	&	$232^{+153}_{-161}$	&	$0.92^{+0.23}_{-0.27}$	&	&	\\
\hspace{0.1em}\vspace{-0.5em}\\
\midrule
$N_T$ (Total)$^{\dagger\dagger}$	&	 \multicolumn{5}{c}{$8.27^{+0.90}_{-1.00}\times 10^{11}$~cm$^{-2}$}\\
\bottomrule
\end{tabular}

\begin{minipage}{0.75\textwidth}
	\footnotesize
	\textbf{Note} -- The quoted uncertainties represent 95\% highest posterior density.\\
	$^\dagger$Column density values are highly covariant with the derived source sizes.
	$^{\dagger\dagger}$Total column density derived from combining posterior column densities of each component. The uncertainty corresponds to the 95\% highest joint posterior density.
\end{minipage}
\end{table*}

\newpage

\subsection{2-cyano-CPD}

\begin{figure}[h!]
    \centering
    \includegraphics[width=\textwidth]{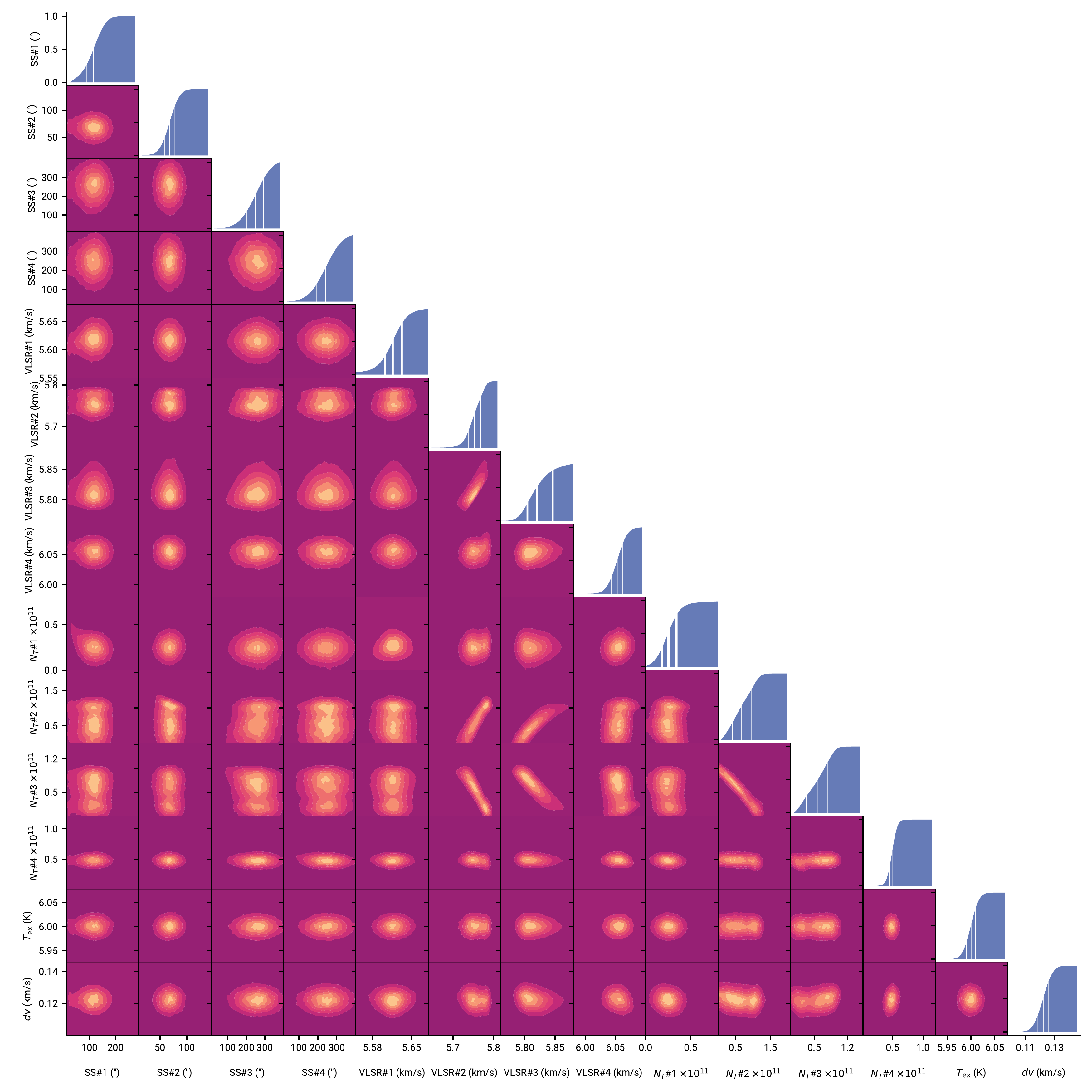}
    \caption{Corner plot for 2-cyano-CPD The diagonal traces correspond to ECDF plots, and off-diagonal plots show the kernel density covariance between model parameters. In the former, lines represent the 25{\nth}, 50{\nth}, and 75{\nth} percentiles respectively. The length scale for the kernel density plots is chosen with Scott's rule.}
    \label{fig:dos-corner}
\end{figure}

Figure \ref{fig:dos-corner} shows the posterior traces for the MCMC modeling analysis of 2-cyano-CPD. These simulations suggest detections in components \#1 and \#4, and components \#2 and \#3 are degenerate with the source sizes effectively poorly constrained in the all cases. Table \ref{table:2cyano} provides summary statistics of the marginalized posteriors.

\begin{table*}[hbt!]
\centering
\caption{2-cyano-CPD best-fit parameters from MCMC analysis.}
\label{table:2cyano}
\begin{tabular}{c c c c c c}
\toprule
\multirow{2}{*}{Component}&	$v_{lsr}$					&	Size					&	\multicolumn{1}{c}{$N_T^\dagger$}					&	$T_{ex}$							&	$\Delta V$		\\
			&	(km s$^{-1}$)				&	($^{\prime\prime}$)		&	\multicolumn{1}{c}{(10$^{11}$ cm$^{-2}$)}		&	(K)								&	(km s$^{-1}$)	\\
\midrule
\hspace{0.1em}\vspace{-0.5em}\\
C1	&	$5.612^{+0.060}_{-0.050}$	&	$110^{+71}_{-90}$	&	$0.27^{+0.21}_{-0.24}$	&	\multirow{6}{*}{$6.00^{+0.03}_{-0.03}$}	&	\multirow{6}{*}{$0.122^{+0.010}_{-0.010}$}\\
\hspace{0.1em}\vspace{-0.5em}\\
C2	&	$5.754^{+0.040}_{-0.039}$	&	$67^{+31}_{-32}$	&	$0.64^{+0.54}_{-0.64}$	&		&	\\
\hspace{0.1em}\vspace{-0.5em}\\
C3	&	$5.836^{+0.126}_{-0.056}$	&	$249^{+140}_{-129}$	&	$0.54^{+0.44}_{-0.51}$	&		&	\\
\hspace{0.1em}\vspace{-0.5em}\\
C4	&	$6.054^{+0.028}_{-0.028}$	&	$239^{+140}_{-135}$	&	$0.48^{+0.15}_{-0.14}$	&		&	\\
\hspace{0.1em}\vspace{-0.5em}\\
\midrule
$N_T$ (Total)$^{\dagger\dagger}$	&	 \multicolumn{5}{c}{$1.89^{+0.18}_{-0.15}\times 10^{11}$~cm$^{-2}$}\\
\bottomrule
\end{tabular}

\begin{minipage}{0.75\textwidth}
	\footnotesize
	\textbf{Note} -- The quoted uncertainties represent the 95\% highest posterior density.\\
	$^\dagger$Column density values are highly covariant with the derived source sizes.
	$^{\dagger\dagger}$Total column density is given as the mean combined posterior column densities of each component. The uncertainty corresponds to the 95\% highest joint posterior density.
\end{minipage}
\end{table*}

\end{document}